\documentclass{article}
\usepackage[utf8]{inputenc}

\usepackage{amsmath}
\usepackage{amsfonts}
\usepackage{enumitem}
\usepackage{amssymb}
\usepackage{setspace}
\usepackage{makeidx}
\usepackage{graphicx}
\usepackage[unicode=true,breaklinks=true,hidelinks]{hyperref}
\usepackage{csquotes}
\usepackage{setspace}
\usepackage{color}
\usepackage{epigraph}

\usepackage{footmisc}
\usepackage{makeidx} 

\newcommand{\ket}[1]{| #1 \rangle}
\newcommand{\bra}[1]{\langle #1 |}
\graphicspath{ {images/} }
\setlist[enumerate]{label*=\arabic*.}
\makeindex
\date{\vspace{-5ex}}

\title{Non-Accessible Mass and the Ontology of GRW}
\author{Cristian Mariani\\
\textit{penultimate version}\footnote{To appear in \textit{Studies in History and Philosophy of Science}, 2022, 91: 270-279.}}

\begin{document}

\maketitle

\epigraph{\textit{This mathematics would allow electrons to enjoy the cloudiness of waves, while allowing tables and chairs, and ourselves, and black marks on photographs, to be rather definitely in one place rather than another, and to be described in ‘classical terms’.}}{J. S. Bell \cite{Bel87}, 190}

\begin{abstract}
\noindent The Mass Density approach to GRW (GRW$_M$ for short) has been widely discussed in the quantum foundations literature. A crucial feature of GRW$_M$ is the introduction of a \textit{Criterion of Accessibility} for mass, which allows to explain the determinacy of experimental outcomes thus also addressing the tails problem of GRW. However, the \textit{Criterion of Accessibility} leaves the ontological meaning of the non-accessible portion of mass utterly unexplained. In this paper I discuss two viable approaches to non-accessible mass, which I call anti-realist and realist, and will defend the latter. First, I show that the anti-realist approach suffers from various objections. Second, I develop an account of non-accessible mass density states as objectively indeterminate states of affairs.  
\end{abstract}

\bigskip

\subsubsection*{Keywords}
Collapse Theories. Mass Density GRW. Ontic Indeterminacy. Quantum Mechanics. Tails Problem.

\bigskip

\section{Introduction}
Collapse theories such as GRW \cite{Ghi86} are among the most discussed solutions to the measurement problem in quantum mechanics (QM). The core idea of GRW is to modify Schrödinger's dynamical equation of standard QM by adding a stochastic and non-linear element. By doing so, the collapse of the wave function can be described as a physical mechanism without making reference to observers or to experimental apparata. In GRW, collapses happen spontaneously and randomly, with a certain probability rate per unit of time. This is achieved by introducing two constants for the spontaneous localization, one for its accuracy in space ($\alpha=10^{-5}$cm), and one for its frequency in time ($\lambda=10^{-16}$s$^{-1}$).\footnote{These values for $\alpha$ and $\lambda$ were proposed in \cite{Ghi86}. I shall notice that during the years different values have been proposed (e.g. \cite{Adl03}), some of which have been empirically falsified. For a recent discussion, see \cite{Tor18}.} The rate is such that for microscopic systems like nucleons, the collapse of the wave function is incredibly rare, which explains why microscopic superpositions have empirically detectable effects (such as the interference pattern in a double slit experiment). However, the rate is also such that for macroscopic systems made of a large number of highly entangled particles, the collapse is practically certain to occur, and this is why superposition states have no effect at the macroscopic scale.

In a series of paper in the early nineties, it was shown that the original GRW proposal from 1986 suffered from various limitations. First, GRW was unable to account for indistinguishable particles, because it does not provide the symmetric or anti-symmetric statistical rules for microscopic constituents. This problem was addressed in \cite{Pea89} and in \cite{Ghi90} with the so-called \textit{Continuous Spontaneous Localization} model, in which the random jumps of GRW are substituted by a continuous dynamical evolution of the Hilbert space. Second, it became clear that particles with different mass have a different frequency $\lambda$ for collapse (for instance, for electrons the probability is about 2000 times lower). Third, Ghirardi \textit{et al} \cite{Ghi95} came to the understanding that the notion of distance in the Hilbert space cannot be translated without ambiguity into the notion of distance in 3D space.

For all these reasons, it became evident that mass should play a prominent role in the collapse mechanism. This consideration lead to the hypothesis that what GRW is really about---what is \textit{out there in the world} and makes the theory true, as philosophers would say---is the distribution of mass over the universe, a field in 3D space.\footnote{In \cite{All14} it is noticed that the notion of \textit{matter} should be preferred over \textit{mass}: “the matter that we postulate in GRWm and whose density is given by the m function does not ipso facto have any such properties as mass or charge; it can only assume various levels of density” (331-2). Since the distinction between \textit{mass} and \textit{matter} does not play any particular role for the purposes of this paper, I will always speak about \textit{mass} to make things easier.} First suggested by Ghirardi Grassi \& Benatti \cite{Ghi95}, this view is known as the Mass Density approach to GRW (GRW$_M$). GRW$_M$ has been defended in various papers since then (\cite{Ghi11}, \cite{Ghi99}, \cite{Bas01}, \cite{Bas04}), and it is one of the major competitors as the correct ontology of GRW.\footnote{The two main competitor approaches are GRW with a wave function ontology only (GRW$_\emptyset$; \cite{Alb96}, \cite{Lew06}) and GRW with a Flash ontology (GRW$_F$; \cite{Tum06}). A third approach is in \cite{Gao17}. It is also worth mentioning Angelo Bassi's suggestion (personal communication) to take GRW as an \textit{effective theory}, an approach that in many ways would render intrinsically incomplete any discussion on the ontology of this theory. For a recent overview on the ontology of GRW, see \cite{Gao18}. This paper focuses on GRW$_M$ only.} 

Any interpretation of QM has to be equipped with a tool for translating the mathematical formalism into empirically testable states. In standard QM this is done by the Eigenstate-Eigenvalue Link (EEL), which entails a one-to-one mapping from eigenvectors in the Hilbert space onto physical properties experienced in the labs. In the case of GRW, however, it is well-known that EEL is too strict as a link, because the dynamics of the theory never evolves into eigenstates, but only very close to them. When a GRW collapse occurs, the wave function gets multiplied by a Gaussian that localizes the system with accuracy given by $\alpha$. And although a large part of the post-collapse state is localized in a small portion of space, the system is also spread infinitely in both sides of the \textit{tails} of the Gaussian. This is known as \textit{tails problem}, and it is among the most discussed issues in the literature on GRW (for an overview, see \cite{Lew03b}, and \cite{McQ15}). 

To overcome this problem and explain the definiteness of outcomes in GRW$_M$, Ghirardi \textit{et al} \cite{Ghi95} define a \textbf{Criterion of Accessibility} for mass (\textbf{CAM} for short) that is meant to play the role of EEL. The upshot is that \textbf{CAM} explains why experimental outcomes correspond to accessible mass only, and therefore why the tails can be neglected. However, this move leaves the ontological meaning of the non-accessible portion of mass utterly unexplained. And in fact, in later discussions on this proposal two positions have emerged. 

\begin{description}

\item \textbf{Anti-Realism}. The mass is real \textit{iff} it is accessible.

\item \textbf{Realism}. All the mass is real, but only a portion is accessible.

\end{description}

\noindent Clearly \textbf{Anti-Realism} and \textbf{Realism} take \textbf{CAM} very differently. According to the former, \textbf{CAM} tells us what is real according to GRW$_M$, so it has an ontological import. According to the latter, \textbf{CAM} has an epistemic role, and only tells us what we have access to experimentally. Although both positions have been defended in the literature,\footnote{\textbf{Anti-Realism} has been endorsed by Ghirardi \textit{et al} \cite{Ghi95}. However, in later writings \cite{Ghi11} Ghirardi seemed to turn to a more nuanced position, though he never endorsed \textbf{Realism} explicitly. An analysis of the differences between the two approaches is hinted at in \cite{Cli00}, and \cite{Myr18}, while a defense of \textbf{Realism} can be found in \cite{Mon04}. As it will emerge in this paper though, the crucial philosophical implications of \textbf{Realism} are not recognized by the authors I have just mentioned. And indeed, even in the most recent literature on GRW (e.g. \cite{McQ15}) there still persists a certain confusion on the correct understanding of \textbf{Realism}. See discussion in \S4.3 of this paper.} I believe that a full-fledged analysis of this issue is still missing. In particular, I contend that the ontological implications of accepting \textbf{Realism} have not yet been fully recognized.  

The aim of this paper is to fill this gap. I will provide several objections to \textbf{Anti-Realism}, and will put forward a novel way of understanding the ontology of GRW$_M$ given \textbf{Realism}. In \S2 I introduce \textbf{CAM} in more details, and show how it promises to overcome the tails problem. In \S3 I give six objections to \textbf{Anti-Realism}, only some of which are already in the literature. In \S4 I propose a novel way of understanding the ontology of GRW$_M$ that is explicitly a form of \textbf{Realism} towards non-accessible mass. I will do so in terms of the distinction between \textit{determinate} and \textit{indeterminate} state of affairs, which I borrow from Wilson \cite{Wil13}. This will help me defend \textbf{Realism} against some recent objections from McQueen \cite{McQ15}. In \S5 I consider some crucial conceptual implications of \textbf{Realism} with respect to the current philosophical debate on GRW$_M$, and in \S6 I conclude.

\section{The Criterion of Accessibility of Mass}
I start in \S2.1 by defining the Mass Density function $\mathcal{M}(\textbf{r},\textit{t})$ of GRW$_M$, which is what gives the ontology of the theory in 3D space. In \S2.2 I introduce the \textbf{Criterion of Accessibility} of mass (CAM), and give a concrete example of how it is to be applied. In \S2.3 I discuss how \textbf{CAM} is supposed to overcome the \textit{tails problem}, and indicate why it is ambiguous between \textbf{Anti-Realism} and \textbf{Realism} towards non-accessible mass.

\subsection{The Mass Density Function}
The crucial conceptual amendment of GRW$_M$ with respect to previous versions of the theory concerns the introduction of a new operator $M(\textbf{r})$ for the Mass Density, which Ghirardi \textit{et al} \cite{Ghi95} define as follows:

\begin{align}
    M(\textbf{r})=\sum_{k}m_{k}N_{k}(\textbf{r})
\end{align}

\noindent Where $k$ are the particles of a given type, $\textbf{r}$ stands for a given spacetime point, and $N$ is the operator describing the number of particles, which is in turn defined as:

\begin{align}
    N(\textbf{r})=a^{\dagger}(\textbf{r})a(\textbf{r})
\end{align}

\noindent In \cite{Ghi86}, the eigenbasis of $N(\textbf{r})$ was the preferred basis in the Hilbert space on which collapses occur. In GRW$_M$, instead, the preferred basis is $M(\textbf{r})$. Consequently, the fundamental dynamical equation of GRW$_M$ is the following:

\begin{align}
\frac{d}{dt}\ket{\psi(t)}=\left[-\frac{i}{\hbar}H+\int d^{3}r M(\textbf{r})V(\textbf{r},t)-\frac{\gamma}{m^{2}_0}\int d^{3}r M^{2}(\textbf{r})\right]\ket{\psi(t)}
\end{align}

\noindent Equation (3) is a Stratonovich dynamical stochastic equation that appears in any GRW theory to modify the Schrödinger's evolution. $V$ indicates the characteristic white noise of the theory, i.e. its inherent stochastic nature, while $\gamma$ encodes the constants for collapse, namely $\lambda$ for the frequency, and $\alpha$ for the accuracy. The only difference with respect to previous versions of GRW, is that (3) acts on the Mass Density operator $M(\textbf{r})$ introduced in equation (1).\footnote{I shall note that eq. (3) is the \textit{Continuous Spontaneous Collapse} (CSL) version of the GRW$_M$ dynamical equation. In a way, we could call it CSL$_M$. I shall stress however, that despite the mathematical differences between the two theories, from an ontological and broadly philosophical perspective there is no significant distinction to be worried about here.} 

The importance of this amendment is that it allows to indicate what the theory is about, its ontology, by defining a Mass Density Function $\mathcal{M}(\textbf{r},\textit{t})$ on 3D space. Let us start by considering a physical system \textit{S} of \textit{N} particles with corresponding Hilbert space $\mathcal{H}(\textit{S})$ of 3\textit{N} dimensions. We then define $\mathcal{M}(\textbf{r},\textit{t})$ as follows:\footnote{What follows is a simplified version of the proofs in \cite{Ghi95}. Although I follow their exposition, I am going to leave some of the technicalities aside.}

\begin{align}
\mathcal{M}(\textbf{r},\textit{t})=\bra{\psi(t)}M(\textbf{r})\ket{\psi(\textit{t})}
\end{align}

\noindent $\ket{\psi(\textit{t})}$ is the normalized vector\footnote{Notice that the Stratonovich equation (equation (3)) does not actually generate normalized vectors. I am going to set this complication aside here, since \cite{Ghi95} provides a way to normalize the vectors.} describing \textit{S} at time \textit{t}, and $M(\textbf{r})$ is the mass density operator defined in equation (1) above. The mass density function defined in (4) provides a mapping of $\mathcal{H}(\textit{S})$ onto the space of functions of \textbf{r} defined in 3 dimensions, at a given time \textit{t}. If we suppose that the physical system $S$ under consideration is the whole universe, and that $\mathcal{H}(\textit{S})$ is its corresponding Hilbert space, then it would follow that equation (4) gives the whole distribution of mass throughout the 3D space. 

\subsection{The Criterion of Accessibility}

$\mathcal{M}(\textbf{r},\textit{t})$ is a many to one mapping, as Ghirardi \textit{et al} \cite{Ghi95} immediately notice. To see this, consider a large number of particles \textit{N} and two regions \textit{A} and \textit{B} both of spherical shape and of the same size, and then compare the following two states $\ket{\psi^{\oplus}}$ and $\ket{\psi^{\otimes}}$: 

\begin{align}
\ket{\psi^{\oplus}}=\frac{1}{\sqrt{2}}\left[\ket{\psi_{N}^{A}}+\ket{\psi_{N}^{B}}\right]
\end{align}

\begin{align}
\ket{\psi^{\otimes}}=\ket{\phi_{N/2}^{A}}\otimes\ket{\phi_{N/2}^{B}}
\end{align}

\noindent Equation (5) expresses a linear superposition of equal amplitudes of the states $\ket{\psi_{N}^{A}}$ and $\ket{\psi_{N}^{B}}$. Equation (6), on the other hand, expresses the tensor product of the states $\ket{\phi_{N/2}^{A}}$ and $\ket{\phi_{N/2}^{B}}$ describing the physical situation of \textit{N}/2 particles in region \textit{A} and \textit{N}/2 particles in region \textit{B}.

Now notice that the states $\ket{\psi^{\oplus}}$ and $\ket{\psi^{\otimes}}$ give rise to the same mass density function $\mathcal{M}(\textbf{r},\textit{t})$ for each region A and B. Consider for example region A:

\begin{align}
\mathcal{M}_{(\textbf{r},\textit{t})}^{\oplus}=\bra{\psi_{t}^{\oplus}}M(\textbf{r})\ket{\psi_{t}^{\oplus}} \approx \frac{1}{2} \bra{\psi_{N}^{A}}M(\textbf{r})\ket{\psi_{N}^{A}} \approx \frac{Nm}{2}
\end{align}

\begin{align}
\mathcal{M}_{(\textbf{r},\textit{t})}^{\otimes}=\bra{\phi_{t}^{\otimes}}M(\textbf{r})\ket{\phi_{t}^{\otimes}} \approx \bra{\phi_{N/2}^{A}}M(\textbf{r})\ket{\phi_{N/2}^{A}} \approx \frac{Nm}{2}
\end{align}

\noindent The same goes for region B. Although the functions generated by (7) and (8) are the same, it is important to discriminate between the states that originate them. Indeed, it is easy to imagine a physical situation where the state $\ket{\psi^{\oplus}}$ behaves like a linear superposition, whereas the state $\ket{\psi^{\otimes}}$ gives rise to a determinate outcome.

To explain the difference between the two states, Ghirardi \textit{et al} \cite{Ghi95} define a criterion for individuating what are the states that give rise to detectable mass distributions (like $\ket{\psi^{\otimes}}$), and what are the states that do not ($\ket{\psi^{\oplus}}$). Their method is simply to define the ratio between the mean expectation value for a given outcome and the variance. We first define the variance $\mathcal{V}(\textbf{r},\textit{t})$ for the mass density operator $M(\textbf{r})$ as follows: 

\begin{align}
\mathcal{V}(\textbf{r},\textit{t})=\bra{\psi(t)}\left[M(\textbf{r})-\bra{\psi(t)}M(\textbf{r})\ket{\psi(\textit{t})}\right]^2\ket{\psi(t)}
\end{align}

\noindent Given $\mathcal{V}(\textbf{r},\textit{t})$, we can define the ratio:

\begin{align}
\mathcal{R}^2(\textbf{r},\textit{t})=\mathcal{V}(\textbf{r},\textit{t})/\mathcal{M}^{2}(\textbf{r},\textit{t})
\end{align}

\noindent Now, if $\mathcal{R}$ turns out to be much smaller than 1, this suggests that the corresponding mass density can be considered detectable, and we call it \textbf{Accessible}. If instead $\mathcal{R}$ is close to 1, the corresponding mass is \textbf{Non-Accessible}. Thus, we can now state the \textbf{Criterion of Accessibility} as follows:

\begin{description}
    \item \textbf{CAM} --- $\mathcal{M}(\textbf{r},\textit{t})$ is accessible \textit{iff} $\mathcal{R}(\textbf{r},\textit{t})\ll1$.
\end{description}

\noindent Given \textbf{CAM}, it can be shown that in the above example the mass corresponding to the state $\ket{\psi^{\otimes}}$ is \textbf{Accessible} because the value of $\mathcal{R}$ is much smaller than 1:

\begin{align}
\mathcal{R}^{\otimes}(\textbf{r},\textit{t})\ll1 
\end{align}

\noindent Contrariwise, for $\ket{\psi^{\oplus}}$ the value of $\mathcal{R}$ is close to 1, and therefore the corresponding mass is \textbf{Non-Accessible}. 

\begin{align}
\mathcal{R}^{\oplus}(\textbf{r},\textit{t})\approx1 
\end{align}

\noindent The \textbf{Criterion of Accessibility}, along with the distinction between \textbf{Accessible Mass} and \textbf{Non-Accessible Mass}, is what explains why, as we should have expected all along, macroscopic superpositions like (5) are not detectable according to the model.

\subsection{Overcoming the Tails Problem: Realism and Anti-Realism}

A generic feature of any collapse theory is that they strive to explain how the mathematical formalism maps onto properties of the physical world. In standard quantum mechanics such a mapping is unambiguously given by the following link:\footnote{See \cite{Gil16} and \cite{Wal16} for critical assessments of the status of EEL in standard quantum mechanics.}

\begin{description}
\item \textbf{Eigenstate-Eigenvalue Link} (EEL). A physical system $s$ has a definite value $v$ of an observable $\mathcal{O}$ \textit{iff} the state of $s$ is an eigenstate of $\mathcal{O}$.
\end{description}

\noindent The result of applying EEL to collapse theories is that no physical system would have any definite property, because the dynamical evolution of these theories never evolves into eigenstates but only close to them. Lacking definite properties, how are we to explain the definiteness of experimental outcomes? This worry was first expressed by Shimony \cite{Shi90}, and then made more explicit by Albert \& Loewer \cite{Alb90} under the name of \textbf{Tails Problem}.

\begin{quote}
        Our worry is that GRW collapses almost never produce definite outcomes even when outcomes are recorded in distinct positions of macroscopically many particles. The reason is that a GRW jump does not literally produce a collapse into an eigenstate of position. A GRW collapse yields one of the states with tails in which almost all the amplitude is concentrated in the region around one of the two components but there is nonzero, though very small, amplitude associated with other regions. ... This means that the post collapse state is not an eigenstate of position and so does not actually assign a definite position to the pointer. (\cite{Alb90}: 284)
\end{quote}
    
\noindent If the post-collapse state is not an eigenstate of position (that is a fact of collapse theories), and if after a measurement the pointer has a definite position (that is an empirical fact), then the post-collapse state cannot explain the definiteness of outcomes. This entails, according to Albert \& Loewer \cite{Alb90}, that collapse theories are not satisfactory solutions to the measurement problem. Notice also that GRW$_M$ makes no exception in this respect, since the dynamics given by equation (3) will never evolve into eigenstates of mass density. 

The existence of the tails in collapse theories also generates a further problem, first individuated by Lewis \cite{Lew97}, that is known as the \textbf{Counting Anomaly} (see also \cite{Lew03a}). Consider a macroscopic marble and a large box, and then take the two states $\ket{\text{in}}$ and  $\ket{\text{out}}$ corresponding to the marble being inside or outside the box respectively. According to Lewis, if we start with the state $\frac{1}{\sqrt{2}}=(\ket{\text{in}}+\ket{\text{out}})$, via the GRW dynamics we almost immediately get either $a\ket{\text{in}}+b\ket{\text{out}}$ or $b\ket{\text{in}}+a\ket{\text{out}}$, with $1>|a|^2\gg|b|^2>0$. Lewis then takes a system of \textit{n} marbles each in a state like $a\ket{\text{in}}+b\ket{\text{out}}$. The \textbf{Counting Anomaly} emerges from considering the state of all the marbles together. The probability of finding all the marbles in the box decreases with the increase of \textit{n}. 

In the intentions of Ghirardi and collaborators, the \textbf{Criterion of Accessibility} for Mass in GRW$_M$ is apt to address the above problems of collapse theories. First recall that the aim of GRW$_M$ is to define, through the mass density function $\mathcal{M}(\textbf{r},\textit{t})$, the distribution of mass all over the universe. Therefore, since it is what is ultimately \textit{out there}, it is natural to assume that the mass distribution is what grounds the truth of empirical predictions. However, we saw in the previous section that not every solution to $\mathcal{M}(\textbf{r},\textit{t})$ refer to a definite state. What \textbf{CAM} does is to tell us what are the states that ground definite empirical predictions. In other words, as it is clear in this passage from Ghirardi, \textbf{CAM} substitutes the \textbf{EEL}:

\begin{quote}
    A property corresponding to a value (or range of values) of a certain variable in a given theory is \textit{objectively possessed or accessible} when, according to the predictions of that theory, experiments (or physical processes) yielding reliable information about the variable would, if performed (or taking place), give an outcome corresponding to the claimed value. Thus the crucial feature characterizing accessibility (as far as statements of individual systems is concerned) is the matching of the claims and the outcomes of physical processes testing the claims (\cite{Ghi97}: 227, italics mine).
\end{quote}

\noindent \textbf{CAM} ensures that the tails can be neglected, for instance in cases such as the \textbf{Counting Anomaly}. For each marble, the only accessible portion of mass is inside the box, and therefore increasing the number of marbles whose accessible mass is inside the box would not change the fact that the tails are not accessible. No marble ‘can be found' outside the box, because there is no accessible mass outside the box. Of course, such an explanation does not say much about \textit{what} the tails are, because once again the status of \textbf{CAM} is left unspecified. In effect, the above passage indicates the ambiguity regarding the meaning of \textbf{CAM}, since Ghirardi uses ‘objectively possessed' and ‘accessible' as if they were synonyms, while it is clear that they evoke two very distinct meanings. Does the fact that the tails are not accessible indicate that they are not \textit{objective}? In the passage above, as well as in other writings (\cite{Ghi99}, \cite{Bas01}, \cite{Bas04}), Ghirardi's attitude has been ambiguous.\footnote{For instance, as correctly pointed out by Clifton \& Monton \cite{Cli00}: “Elsewhere, Ghirardi and Grassi (\cite{Ghi96}, p. 376) have written, with regard to the term ‘objective’, that ‘both usual meanings of that term (i.e. “real” or “opposite to subjective”) do not fit with the sense which emerges for it from our work’"(8).} 

We can distinguish two approaches to this issue.\footnote{Monton \cite{Mon04} has been the first who recognized that we can take two distinct positions towards \textbf{CAM}. He calls them \textit{Accessible Mass Density Link} and \textit{Mass Density Simpliciter Link}, and argues for the latter. Although my position is sympathetic to Monton's, I shall stress in section \S4.3.2 why his view is not developed enough, and therefore is the target to several objections.} First, we could stress that \textbf{CAM} has an ontological import, i.e. it tells us what is real and what is not. The consequence of this approach is that non-accessible mass states, along with the tails of the wave function, are not real:

\begin{description}

\item \textbf{Anti-Realism}. The only solutions to $\mathcal{M}(\textbf{r},\textit{t})$ that correspond to physical states are those that are accessible according to \textbf{CAM}.

\end{description}

\noindent On the contrary, if we put the accent on an epistemological understanding of \textbf{CAM}, we have: 
\begin{description}

\item \textbf{Realism}. Every solution to $\mathcal{M}(\textbf{r},\textit{t})$ correspond to a physical state, but only some of these are accessible---namely, those given by \textbf{CAM}.

\end{description}

\noindent \textbf{Realism} retains the explanation of the definiteness of experimental outcomes given by \textbf{CAM}, but does not further assume that states that are not accessible are not real or objective. 

Notice now that the two views face very distinct explanatory challenges. On the one hand, \textbf{Anti-Realism} has the difficult task of explaining why, if what is real is tantamount to what is accessible to us, GRW$_M$ is after all a substantive improvement with respect to standard quantum mechanics. On the other hand, \textbf{Realism} has to come up with an explanation of the physical meaning of non-accessible mass states. Why, if they are physical, should they not be accessible? What does the distinction between the two kinds of mass, accessible and non, ultimately amounts to? As I shall make clear in \S4, such an explanation cannot, on pain of circularity, be based on the distinction between what is accessible and what is not. The distinction has to be ontological. 

In \S4 of this paper I will argue that \textbf{Realism} can be coherently defended. Before this, however, I shall first review in \S3 the main motivations for rejecting \textbf{Anti-Realism}. 

\section{Six Objections Against Anti-Realism}
While there is no explicit defense of \textbf{Anti-Realism} in the literature, it is worth considering this view for at least two reasons. First, as already mentioned, Ghirardi and collaborators have sometimes suggested a view close to this one. Second, \textbf{Anti-Realism} has already been explicitly rejected by some authors. A review of these arguments, were they to be successful, will provide indirect support to \textbf{Realism}. To my knowledge, we can find four distinct objections in the literature, in \cite{Mon04}, \cite{Tum11}, \cite{McQ15}, and \cite{Lew16}. After considering them respectively in \S3.1, \S3.2, \S3.3, and \S3.4, I will provide two new objections in \S3.5 and \S3.6.

\subsection{Objection 1: The Disappearance and Reappearance of Particles}
Monton \cite{Mon04} considers a physical exemplification of the states from equations (5) and (6), namely $\ket{\psi^{\otimes}}$ and $\ket{\psi^{\oplus}}$. Suppose we send a test particle between region A and B. Given gravitational effects, if A and B are in the state $\ket{\psi^{\otimes}}$, the test particle would continue its trajectory without being deflected. If, instead, A and B are in the state $\ket{\psi^{\oplus}}$, the test particle would be deflected either towards region A or towards region B with equal probability. The consequence, according to Monton, is that according to \textbf{Anti-Realism}:

\begin{quote}
    [...] the test particle with inaccessible mass is nowhere in the universe: since its mass density is not accessible, it is not real. For a microscopic test particle, its mass could be inaccessible for a long time. In fact [...] objects would often be popping out of and into existence, as the accessibility of their mass changed. While I do not have a knock-down argument as to why this is unacceptable, I maintain that this is a serious anomaly. I admit that the evolution of systems according to quantum mechanics is non-classical, but the regular disappearance and reappearance of particles, where sometimes the disappearance is for extended periods of time, moves beyond the realm of the benignly non-classical and into the realm of the anomalous. (\cite{Mon04} 14-15)
\end{quote}

\noindent I agree here with Monton that this argument is not a knockdown objection. However, I think it is more pressing once we consider that Bassi and Ghirardi (\cite{Bas04}: 90) claim that GRW \enquote{does not contemplate creation and annihilation of particles}. If this is the case, then there seems to be little room for \textbf{Anti-Realism} to explain the transition from non-accessible to accessible mass density states. 

\subsection{Objection 2: The Role of Observers I}
A different objection can be found in Tumulka \cite{Tum11}. The aim of Tumulka is to focus on whether, if \textbf{Anti-Realism} is adopted, GRW$_M$ can be considered a viable candidate as an interpretation of quantum mechanics. He focuses on the tails problem and the solution to it given by \textbf{CAM}, and argues: 

\begin{quote}
    [The tails problem] concerns whether GRW theories provide a picture of reality that conforms with our everyday intuition. Such a worry cannot be answered by pointing out what an observer can or cannot measure. Instead, I think, the answer can only lie in what the ontology \textit{is like}, not in what observers see of it. (\cite{Tum11}: 9)
\end{quote}

\noindent This is, I believe, the strongest argument against \textbf{Anti-Realism}. The whole project of GRW is to provide an explanation of measurement processes without making reference to observers. The tails problem threatens the viability of this project. As a matter of fact, although neither \textbf{CAM} nor \textbf{Anti-Realism} mention observers directly,\footnote{I thank Peter Lewis for making this point.} the very idea of \textit{accessibility} is quite naturally related to the epistemic capacity of observers. Without observers involved, there would be no reason to appeal to this notion. Thus, if we solve the tails problem by accepting \textbf{Anti-Realism}, it seems that we are in fact, once again, addressing the measurement problem by referring to observers.

\subsection{Objection 3: The Role of Observers II}
In his \cite{McQ15}, McQueen provides an exhaustive assessment of the tails problem in GRW, and presents four distinct versions of it. Since the third and fourth largely depend on the second one, the crucial focus is on the distinction between the first and the second, which he calls \textit{The Bare Tails Problem} and \textit{The Structured Tails Problem} respectively. The \textbf{Bare} one is what we met in the previous section---the existence of the tails entails the need to revise the EEL. According to McQueen this version of the problem is not particularly pressing, since one could simply revise the EEL and maintain that there is some vagueness regarding whether a property supervenes on the quantum state or not---e.g. by allowing for properties to possess definite values even when the corresponding quantum state is sufficiently close to an eigenstate.\footnote{A proposal introduced in \cite{Alb96} and named \textit{Fuzzy Link}.}

While the \textit{Bare} version can be solved, according to McQueen the more severe version is the \textit{Structured} one, first recognized in \cite{Cor99}, and then further developed in \cite{Wal08}. As McQueen defines it:

\begin{quote}
    \textbf{The Structured Tails Problem}: If the collapse centre structure determines a particle configuration, then so do the structures in the tails. This is because the tails and the collapse centre are structurally isomorphic (or at least relevantly structurally similar). Nothing about low mod-square value can suppress this isomorphic structure. The consequence is an Everettian many-worlds ontology. (\cite{McQ15}: 7)
\end{quote}

\noindent According to McQueen, \textit{The Structured Tails Problem} can only be solved if we find a way to break the symmetry between the low-density and the high-density mass configurations. Without breaking the symmetry, the consequence would be that the ontology of GRW is “Everettian in disguise". McQueen reviews various proposals for breaking the symmetry, and among them he also considers Ghirardi \textit{et al} view \cite{Ghi95}, which he explicitly takes as a form of what I call \textbf{Anti-Realism}.

\begin{quote}
    They define the low-density matter in the tails as “inaccessible” (i.e. observers cannot directly measure it) and so “not objective”. (\cite{McQ15}: 9)
\end{quote}

\noindent As I have shown in \S2, the inference from \textit{inaccessible} to \textit{not objective} in GRW$_M$ can be resisted. And in fact, McQueen himself elsewhere in his paper considers this option. While I will come back to this in \S4.3.1, for now I shall focus on McQueen's motivation for rejecting \textbf{Anti-Realism}, which is the following:

\begin{quote}
     Ghirardi et al. cannot (without circularity) appeal to observers \textit{until} they've solved the structured tails problem. After all, the observers in the tails can access the matter in the tails. So what's accessible to observers can only be defined by this theory after the structured tails problem has been solved. The fact that the observers in the collapse centre cannot access the matter-density in the tails does not appear to speak to the real tails problem. (\cite{McQ15}: 10)
\end{quote}

\noindent McQueen's point is close in spirit to Tumulka's: the solution to the tails problem cannot appeal to observers; \textbf{Anti-Realism} solves the tails problem by appealing to what we can and cannot observe; therefore \textbf{Anti-Realism} is not a solution to the tails problem. Notwithstanding the analogy, two important remarks are in order. 

First, McQueen is more explicit than Tumulka, I think correctly, that the problem emerges from the \textit{circularity} of the explanation given by \textbf{Anti-Realism}. Tumulka simply claims that the explanation should not appeal to observers, while McQueen is more explicit as to why this cannot be the case: the explanation would be circular. 

Second, notice that McQueen's conclusion is clearly stronger than Tumulka's in that it also adds the “Everett in disguise" morale, which stems from the way he defines the structured tails problem, following \cite{Cor99}---see also \cite{Wal08} and \cite{Vai14}. I will come to this issue later on, when I will consider McQueen's objections to my own proposal (see \S4.3). 

\subsection{Objection 4: The Role of the Quantum State}
A further objection, somewhat also close in spirit to Tumulka's, is to stress that the original GRW was meant to consider physical properties as fully described by the quantum state (see \cite{Bas04}: 103-4).\footnote{I notice that Bassi \& Ghirardi (\cite{Bas04}: 104) speak about the departure from the view that the quantum state fully describes properties as a price that is worth paying in order to address the counting anomaly. However, they make this claim precisely because they are not considering \textbf{Realism} towards non-accessible mass as an option.} Lewis alleges something close to this:

    \begin{quote}
    Massy GRW [...] eschews any direct link between the quantum state and physical properties. The condition for a marble being in a box makes no mention of the quantum state: The marble is in the box if and only if the associated region of high mass density is in the appropriate region of space. (\cite{Lew16}: 94)
    \end{quote}

\noindent The departure from the view that the quantum state directly describes physical properties is not a problem \textit{per se}. Nonetheless, such a departure is a cost, and \textit{all other things being equal} we should try and avoid it. This gives further indirect support to \textbf{Realism} towards non-accessible mass.

\subsection{Objection 5: Mathematical Artifacts}
One could argue that according to \textbf{Anti-Realism}, the mass density corresponding to states like $\ket{\psi^{\oplus}}$ (eq. 5) simply does not represent anything physical. It is, in a way, just a mathematical artifact of the theory that we should not take seriously as far as the ontology is concerned, and yet it is useful to make sense of our observations.\footnote{I thank Vincent Lam for inviting me to consider this view on behalf of \textbf{Anti-Realism}, and Giuliano Torrengo for useful discussions on this part of the paper.} 

Indeed, the idea that a subset of the solutions to a fundamental equation does not correspond to physical states is not something new. To mention but one example, a similar idea has been advocated in the context of the explanation of quantum statistics (see e.g. \cite{Fre06}). Since fundamental particles do not obey classical statistics, it is well-known that we strive to provide identity conditions for them. One proposal is to advocate a notion of \textit{primitive identity} and reject Leibniz's principle of identity of indiscernibles. By doing so, however, the consequence is that a large part of the possible statistical arrangements, although possible \textit{in principle}, can never be experienced in practice. In this sense, it has been suggested that the states we do not experience are just mathematical artifacts. However, such a conclusion has been explicitly challenged (see \cite{Red91}) on methodological grounds: the less surplus structure the better (for discussions see \cite{Bel00}).

One could protest that the above example is not analogous to the case I am discussing. While in the case of quantum statistics it seems fair to say that the primitive identity of particles does not play any role in explaining physical phenomena,\footnote{Which, of course, does not entail that they play \textit{no} role whatsoever.} states like $\ket{\psi^{\oplus}}$ have a physical meaning, since they give us information about possible or future measurements.\footnote{Thanks to Peter Lewis here.} To be surplus structure does not entail to be a mathematical artifact. However, I contend that in both cases, and virtually whenever part of the theory does not have a corresponding ontology, we should ask ourselves \textit{why} this happens and look for an explanation. The explanation cannot be \textit{ad hoc} or arbitrary, while it could of course be justified for its theoretical virtues. In the case of \textbf{Non-Accessible} mass, it seems highly arbitrary to claim that the only states that exist according to the theory are those that we can have access to. Also because, as I am about to show with the next and final objection to \textbf{Anti-Realism}, the very definition of \textbf{CAM} is inherently arbitrary.   

\subsection{Objection 6: Arbitrariness}
The sixth objection to \textbf{Anti-Realism} stems from considering the inherent pragmatic nature of the \textbf{Criterion of Accessibility}.\footnote{In \cite{Lew03b}, Lewis argues at length against Albert \& Loewer's \textit{fuzzy link} \cite{Alb96}. As a reviewer of this journal points out, there may be a similarity between this critique and my objection to \textbf{Anti-Realism} based on \textit{arbitrariness}. Both Lewis' criticism and the one I presented are meant to show that the link cannot be a matter of stipulation. However, the two objections disagree as to why it is so. According to Lewis, the exact form of the link cannot be \enquote{a matter of stipulation}, because \enquote{the actual form of the link may rule out spontaneous collapse theories altogether} (\cite{Lew03b}: 1444). According to my objection, instead, \textbf{CAM} cannot have an ontological import because this would entail that what exists is a matter of stipulation. I thank the reviewer for inviting me to spend a few words on this.} We can imagine revising the criterion slightly as follows:

\begin{description}
    \item \textbf{CAM*} --- $\mathcal{M}(\textbf{r},\textit{t})$ is accessible \textit{iff} $\mathcal{R}(\textbf{r},\textit{t})\ll\eta$.
\end{description}

\noindent Where $\eta\approx1$. The difference between \textbf{CAM} and \textbf{CAM*} would clearly be negligible, since they would agree on virtually any physical state as to whether the corresponding mass is accessible or not. And if this is true, then of course it would also be true for any revision of \textbf{CAM} where we have another number instead of $\eta$, say $\mu$, different from $\eta$ but still such that $\mu\approx1$. 

Of course, all these complications are just unnecessary since \textbf{CAM} in its most simplified version already does the job. Although this is certainly true, what the above example shows is that it is at least in principle possible to revise \textbf{CAM} in an infinite number of ways, which shows its inherently arbitrary nature once taken with an ontological meaning. Physics is full of examples where the search for a more and more limited margin of error produces more accurate results. For instance, for the Atlas experiment at CERN, the margin of error for experiments aimed at the discovery of new particles such as the Higgs Boson is fixed by providing five different parameters \cite{CERN}. Experimental physicists might obviously discuss whether this margin of error is enough to take seriously the experimental results, as well as they could discuss how it could be improved. But of course, the fact that the margin of error might change, or could be made smaller through technological improvements does not show that \textit{the ontology} itself has changed accordingly! Now, if \textbf{CAM} is really supposed to have an ontological import, it would follow that the way in which we define what is real according to the theory is completely arbitrary: depending on whether we choose \textbf{CAM}, or \textbf{CAM*}, and so on, for instance, we would get different answers as to what exists according to the theory. 

\section{The Realist Approach to Non-Accessible Mass}

As I suggested at the end of \S2, the crucial explanatory challenge for \textbf{Realism} consists in showing what is the physical difference between the states that give rise to accessible mass density distribution (such as $\ket{\psi^{\otimes}}$ of equation (6)) and the states that do not (like $\ket{\psi^{\oplus}}$ of equation (5)). I agree with Tumulka that the answer to this ‘‘can only lie in what the ontology \textit{is like}, not in what observers see of it" (\cite{Tum11}: 9). In this section I will propose a way of addressing this issue by exploiting some tools from analytic metaphysics in order to understand the difference between the two kinds of mass.\footnote{Since I am aware that not everyone will share the methodology I employ here, I cannot but assume that metaphysics can sometimes be of good use to address problems in the philosophy of physics. This methodological approach has been recently defended in various works (\cite{Fre12}, \cite{Mor13}, \textit{inter alia}) aiming to address the scepticism towards analytic metaphysics, and suggesting that pure \textit{a priori} reasoning can be put into service as a sort of ‘toolbox' for philosophy of science. Although I will not explicitly defend this approach, I hope that the reader who disagrees with it might eventually appreciate its utility once applied to concrete cases, as it is in this paper.}

The guiding idea of my proposal is that we understand non-accessible mass states as \textit{objectively indeterminate states of affairs}.\footnote{There may be ways of defending \textbf{Realism} that differ from mine, and perhaps even that do without indeterminacy. For instance, a reviewer of this journal proposes that, although states that give rise to an accessible mass and states that give rise to a non-accessible mass have no ontological difference, they produce different \textit{physical outcomes} because of the distinct type of physical interaction between system and experimental apparatus. The idea would be, roughly, that when the mass is accessible the system-apparatus interaction gives rise to a detection, while it is non-accessible when there is no detection. By accepting a view along these lines, we would not be compelled to accept indeterminacy, or so the thought goes. I grant that this view is interesting and worth developing. I notice however, that the major challenge for \textbf{Realism} is precisely to explain why, if there is no difference between the two states (as this view suggests), the system-apparatus interaction is different, and why it gives rise to a detectable or a non-detectable outcome. If we assume that the only difference is that in one case we have a detection, and in one case we do not, we are in fact assuming what we are trying to explain. I am thankful to the reviewer of this journal for suggesting this proposal and for inviting me to say more about it.} In \S4.1 I introduce Wilson's \cite{Wil13} definition of an indeterminate state of affairs, and I show how this notion is to be applied to quantum mechanics following the work of Calosi \& Wilson \cite{Cal18}. In \S4.2 I apply the notion to GRW$_M$ by introducing what I call the \textbf{Degree Link}, that is a revision of \textbf{EEL} explicitly designed to allow for the existence of indeterminate states of affairs. In \S4.3 I respond to some potential objections. 

\subsection{Indeterminate States of Affairs in Quantum Mechanics}
Ontic or metaphysical indeterminacy (henceforth: OI) is, very roughly, a kind of indeterminacy that is independent both from our knowledge of the world and from the language we use. Philosophers have long been sceptical about the possibility of such a kind of indeterminacy (see \cite{Lew86}, \cite{Eva78}). However, recent years have seen a resurgence of interest in OI. First, it has been shown that we can provide coherent models of OI (for an overview, see \cite{Aki14}), thus addressing the scepticism towards this notion. Second, some authors came to the understanding that physics, quantum mechanics in particular, might give us an instance of OI, along with naturalistic motivations for taking it seriously.

Although there are various proposals on how to understand OI, in this paper I restrict my attention to Wilson's approach \cite{Wil13}. The reason for this limitation is that, as shown in many papers (\cite{Dar10}, \cite{Dar14}, \cite{Sko10}, \cite{Bok14}, \cite{Wol15}, \cite{Cal18}, \cite{Calms}), Wilson's model seems more apt to understand the peculiar kind of OI that we find in quantum mechanics.\footnote{It is fair to mention that not everyone agrees on this. In effect, it has been argued that other models of OI could be applied to quantum mechanics (see especially \cite{Tor17}, \cite{Dar19}, \cite{Mar21b}, and \cite{Fle21}). I will not enter this debate here, since my main interest is to apply the notion of OI, no matter which one, to GRW$_M$. For a recent overview of the debate on quantum indeterminacy, see \cite{Calmar}.} 

Here's Wilson introducing her account:

\begin{quote}
    Determinable-based [OI]: What it is for an SOA [State of Affairs] to be [OI] in a given respect \textit{R} at a time \textit{t} is for the SOA to constitutively involve an object (more generally, entity) \textit{O} such that (i) \textit{O} has a determinable property \textit{P} at \textit{t}, and (ii) for some level \textit{L} of determination of \textit{P}, \textit{O} does not have a unique level-\textit{L} determinate of \textit{P} at \textit{t}. (\cite{Wil13}: 366)
\end{quote}

\noindent According to Wilson, we can give a \textit{reductive} account of OI by allowing for the possibility that objects sometimes lack a unique determinate property of the corresponding determinable. The main theoretical price we have to pay, is the rejection of the plausible idea that determinable properties are nothing but disjunctions of their determinates. This is however a price worth paying, since what we gain is a very clear understanding of what OI really amounts to. This, in turn, allows us to respond to the scepticism towards the notion. 

Wilson's model allows to distinguish between \textit{determinate} and \textit{indeterminate} states of affairs. However, what is more important for us is that this distinction can be applied to quantum mechanics, and Calosi \& Wilson \cite{Cal18} have done precisely so. An indeterminate state of affairs in quantum mechanics can be understood as an object, say an electron \textit{e}, instantiating a certain determinable property, say \textit{spin$_x$}, but no unique corresponding determinate \textit{up} or \textit{down}.\footnote{Of course, we need to grant that the \textit{determinable}/\textit{determinate} distinction can be applied to properties such \textit{spin}. See Wolff \cite{Wol15}, who argues extensively in favor of this idea. Furthermore, one could insist that systems lacking unique determinate properties, such as \textit{up} or \textit{down}, also lack the corresponding determinable (\textit{spin} in the relevant direction). Glick \cite{Gli17} argues for this view, which he calls the \textit{Sparse View}. I will set this option aside in this paper, and simply notice that the lack of determinable properties in Glick's \textit{Sparse View} could also be seen as the presence of some other form of indeterminacy, thus leaving open the issue of whether similar arguments to the one I develop could be put forward. For different arguments against the \textit{Sparse View}, see \cite{Calms}.} Such indeterminate states of affairs are arguably pervasive in quantum mechanics. As argued in \cite{Cal18}, if we take the EEL to be the correct way to ascribe physical properties out of the quantum state, it follows that systems that are not in an eigenstate of a certain observable $\mathcal{O}$ do not have a unique determinate property for $\mathcal{O}$, precisely as required by Wilson's definition of OI. 

The \textit{non uniqueness} requirement for determinate properties can be satisfied in three ways. Let us consider an electron \textit{e}, and one of its observables $\mathcal{O}$, say its spin$_x$. We have the following three possibilities:

\medskip
\begin{description}
\item \textbf{Gappy}. \textit{e} has the determinable spin$_x$ corresponding to the observable $\mathcal{O}$, and it does not have \textit{any} of the determinates of the determinable property spin$_x$ corresponding to $\mathcal{O}$ (\textit{e} is neither \textit{up} nor \textit{down}).\\
\medskip
\item \textbf{Glutty Relativized}. \textit{e} has the determinable spin$_x$ corresponding to the observable $\mathcal{O}$, and it has \textit{more than one} determinate of the determinable property spin$_x$ corresponding to $\mathcal{O}$, each \textit{relatively} to some target (\textit{e} is \textit{up} relative to some target, and \textit{down} relative to some other target).\\
\medskip
\item \textbf{Glutty Degree}. \textit{e} has the determinable spin$_x$ corresponding to the observable $\mathcal{O}$, and it has \textit{more than one} determinate of the determinable property spin$_x$ corresponding to $\mathcal{O}$, each \textit{with a degree} less than 1 (\textit{e} is \textit{up} with degree \textit{d} less than 1, and \textit{down} with degree \textit{d*} less than 1, with \textit{d}+\textit{d*}=1).\\
\end{description}

\noindent In \cite{Wol15} it is argued that \textbf{Gappy} should be preferred, while in \cite{Cal18} we find reasons for adopting \textbf{Glutty Degree}. \textbf{Glutty Relativized} has been considered in \cite{Calfc} in the context of Rovelli's relational interpretation of quantum mechanics.\footnote{For a recent analysis of these options, see \cite{Calggg}.}

The model of quantum indeterminacy I have just sketched is meant to be independent from the specific interpretation of QM one wishes to adopt, as also stressed in \cite{Cal18}. However, we should remind ourselves that the account heavily relies on the EEL to assess which state of affairs is indeterminate, and of course the EEL is rejected by almost every interpretation of the theory. I shall leave this general discussion aside, since my focus is on GRW$_M$ only.\footnote{For discussions on quantum indeterminacy in the different interpretations, see \cite{Cal18}, \cite{Gli17}, \cite{Calfc}, \cite{Calms}.}

\subsection{The Degree Link}
As I have shown throughout this paper, GRW$_M$ needs to revise or substitute the EEL in order to explain the definiteness of outcomes. In Ghirardi's intentions, the \textbf{Criterion of Accessibility} (CAM) is a way of doing so, but it is inherently ambiguous regarding the status of non-accessible mass. The proposal I put forward is a form of \textbf{Realism} which aims to understand states of non-accessible mass as indeterminate states of affairs. Since all the mass exists, both accessible and non, \textbf{Realism} cannot take \textbf{CAM} to be a satisfactory way of revising EEL. My proposal is that we introduce the following link for GRW$_M$:
\begin{description}
    \item \textbf{Degree Link} (DL). A physical system $s$ has a value $v$ of an observable $\mathcal{O}$ \textit{to the extent of} the squared projection of its state onto the eigenstate of $\mathcal{O}$.\footnote{This proposal is inspired by Lewis' \textbf{Vague Link} (\cite{Lew16}: 89), but it also differs from it in a crucial way. Here is Lewis: \begin{quote}
        According to the vague link, my coffee mug almost entirely possesses the determinate property of being on top of my desk, but it also very slightly possesses the determinate property of being inside the drawer. Because the degree of possession of competing properties is so slight, for all practical purposes I can say that the coffee mug is on the desk. (89-90)
    \end{quote} The difference lies in the fact that Lewis recurs to FAPP reasoning to explain why, in his example, the coffee mug is on the desk. On GRW$_M$ we can still appeal to \textbf{CAM} to explain why the mug is on the desk; because all the accessible mass of the mug is on the desk. However, the \textbf{DL} will also tell us that the Mug is also inside the drawer with a certain degree. I am soon going to say more on the interaction between \textbf{CAM} and \textbf{DL}.} 
\end{description}

\noindent According to \textbf{DL}, the ascription of properties comes in degrees given by the quantum state. This entails a widespread indeterminacy in the ontology of GRW$_M$: given the existence of the tails, virtually every property is instantiated with a degree less than 1. However, there is no reason to consider such pervasive indeterminacy in property instantiation as problematic once we consider the possibility to account for it within the framework sketched in \S4.1. And in fact, it is rather straightforward to see that the indeterminate states in GRW$_M$ with the \textbf{DL} are instances of states of affairs of the \textbf{Glutty Degree} kind, on which more than one determinate is jointly instantiated, each with a degree less than 1. 

Now, if every state in GRW$_M$ is indeterminate, one could argue that there would be after all no real difference between accessible and non-accessible mass. In effect, the only difference is a matter of degree, and so it is a quantitative rather than qualitative one. But then the worry could be that we find ourselves back at the start: what explains why we have access only to a portion of the mass, and why do we have determinate outcomes of experiments? This is a fair worry, and one that needs to be addressed in depth. However, I believe a convincing response can be given by stressing that GRW$_M$ with the \textbf{Degree Link} still retains the \textbf{Criterion of Accessibility} for mass, though only \textit{as a pragmatic tool}. Let me unpack this thought.

Let us start by noticing that the state of the pointer after a measurement is not an indeterminate state of affairs. However, we also know that the quantum state that grounds the state of the pointer is indeterminate, given the existence of the tails. How is it possible that a determinate state of affairs (the pointer) is grounded on an indeterminate one (its mass)? The reason is that there has to be a \textit{epistemic threshold} in the degree of instantiation, after which the mass becomes accessible and gives rise to determinate states. The role of \textbf{CAM} is simply to tell us that there exists such a \textit{threshold}, otherwise the definiteness of outcomes would not emerge. But it is a fact that measurements have a unique determinate outcome, and if these outcomes are grounded on the indeterminate ontology of GRW$_M$ with the \textbf{Degree Link}, this means that a \textit{threshold} exists.\footnote{A reviewer of this journal raises an interesting question about how the threshold is supposed to arise from the indeterminate ontology. I think that there may be various options here, and I do not mean to rule out the possibility of a better explanation than the one I give (for instance, of a \textit{mereological} explanation, as the reviewer suggests). The attitude I defend in this paper is however more conservative, and yet I believe this is enough to make my point. By claiming that the threshold is \textit{epistemic}, what I mean is that the most natural explanation of it may be related to our perceptions, for we could safely assume that it would be impossible to perceive indeterminate states of affairs. I thank two anonymous reviewers of this journal for inviting me to say more on this issue.}

Notice also that we may well be ignorant about the exact location of the threshold. It could be, in other words, that \textbf{CAM} is vague in a sense very similar to the one advocated by Williamson \cite{Wil94}, i.e. in an epistemic sense. However, this is not particularly problematic once we have clarified that the role of \textbf{CAM} is not ontological (as it is for \textbf{Anti-Realism}), but it is rather purely pragmatic. 

\subsection{Reply to McQueen}
In a nutshell, the argument for the \textit{threshold} is the following: if (i) there are determinate unique outcomes of experiments, and if (ii) the correct ontology of GRW$_M$ encompasses indeterminate state of affairs as given by the \textbf{Degree Link}, then there has to be a \textit{threshold}. In this part of the paper I am going to consider two potential objections to this argument, one against each of (i) and (ii). Given the novelty of the proposal, there are of course no objections in the literature explicitly against it. However, McQueen \cite{McQ15} alleges something close to the view just proposed, and provides the two objections I am about to consider.

\subsubsection{Objection to Premise (i): The Structured Tails Problem}
According to McQueen the most severe version of the tails problem is the so-called \textit{Structured} one (see \S3.3 of this paper). To recall, this problem can only be solved if we find a way to break the symmetry between the low-density and the high-density mass configurations. Without breaking the symmetry, the ontology of GRW is ‘Everettian in disguise'. Furthermore, my proposal admittedly does not break the symmetry, since as I have argued the distinction between what McQueen calls high and low-density states is not ontological, but only comes in degrees. Thus, one could argue that the proposal suffers from the \textit{Structured Tails Problem}. This conclusion would in turn undermine premise (i) of the argument for the threshold, since it would entail that there are no unique outcomes of experiments.

Although my response to this worry is in fact a quite general reaction to McQueen's Everettian reasoning, it will be useful to further justify the pragmatic approach to the \textbf{Criterion of Accessibility} I advocated. 

At the beginning of his paper, McQueen characterizes the solutions to the measurement problem in terms of Maudlin's \cite{Mau95} three mutually incompatible propositions: (A) the wave function completely specifies all physical properties; (B) the wave function always evolves with a linear dynamics; and (C) measurements always have a single, determinate outcome. The measurement problem can be solved by denying at least one among (A), (B), and (C). Now, theories like GRW deny (B), as shown in \S2.1. It is crucial to notice that once (B) is denied, there is no conceptual need to \textit{also} deny (C); it would simply be useless. And indeed, collapse theories do not deny (C). The reasons not to reject (C) could be of a methodological or common sense kind: we always experience definite outcomes, and denying this by complicating the ontology is something we should avoid if possible. This does not mean of course that one could not find good, independent motivations for rejecting (C). Many-worlds type of interpretations do attempt to provide good motivations in this direction. Nonetheless, it is also clear that \textit{all other things being equal} it is better not to reject (C). So, it should be taken as a virtue of GRW (and, in this respect, of Bohmian mechanics too, which denies (A) but neither (B) nor (C)), that proposition (C) is upheld. 

But once we maintain the truth of (C), the question of what justifies it becomes ill posed. (C) reflects a claim that we prefer our theories to maintain as a sort of \textit{desideratum}. These considerations, I believe, cast some doubts over the conclusions drawn by McQueen from the \textit{Structured Tails Problem}. By accepting an indeterminate mass density with distinct properties each instantiated to a certain degree, we do not need distinct worlds each corresponding to distinct properties. Indeterminacy is a price to pay, but once we are willing to pay this price there is no need to also pay the further price of multiplying worlds. And as a matter of fact, McQueen himself (\cite{McQ15}: 14) considers this as a way out, but dismisses it on grounds that indeterminacy is inconsistent (I will come back to this shortly, in \S4.3.2.). Of course a many-worlds ontology is always an option. Yet, once the measurement problem is solved without rejecting (C), it is not clear why we should take this option seriously. 

Having understood that the many-worlds conclusion is not forced upon us, we also get a better grasp on the justification to the pragmatic approach to \textbf{CAM}. (C) can be assumed within the context of collapse theories, and the argument for the existence of a \textit{threshold} has it as one of its premises. If (i) outcomes of experiments are unique and determinate, and if (ii) the ontology is one of properties instantiated with degrees, then there has to be a \textit{threshold}. 

\subsubsection{Objection to Premise (ii): Indeterminate Existence}
McQueen considers as a possible response to the tails problem the view proposed by Clifton \& Monton \cite{Cli99}, according to which we could postulate a direct connection between \textit{probability of occurring} and \textit{actual existence}. In a way, Clifton \& Monton's view is close to mine for at least two reasons. First, their view is explicitly a form of \textbf{Realism} towards non-accessible mass. Second, and more important, Clifton \& Monton aim to elucidate the ontology of GRW by allowing for a kind of ontic indeterminacy. Their view is however underdeveloped, and therefore target to McQueen's following criticism:

\begin{quote}
Clifton and Monton could postulate that mod-square plays a \textit{further} role beyond what GRW intended. This postulate would relate the existence of what composes macro-objects with mod-square values such that existence comes in degrees and there are borderline cases of whether or not something exists. In GRW$_M$ the mass-density would fade out of existence as its associated value drops below 1-q. And as the associated value goes above 1-q a higher level of existence (as well as density) is exemplified. Such a theory might conceivably be true. But much more work is needed to make sense of the idea of indeterminate existence, and to develop the theory more generally. (\cite{McQ15}: 9)
\end{quote}

\noindent McQueen's argument is that, though conceivable, Clifton \& Monton's view has the difficult task of explaining how existence can come in degrees.\footnote{An example can be found in Smith's \textit{Degree Presentism} \cite{Smi02}, where degrees of existence are ascribed differently depending on the temporal distance from the present moment, with the present being real with a degree = 1.} I believe his argument could be made even stronger by considering that the notion of indeterminate existence could face the same strong objection made by Evans \cite{Eva78} against indeterminate identity. At least, it would for those who believe that ‘to exist' means nothing else than ‘to be identical to something'.\footnote{Thanks to Claudio Calosi here.}

If successful, McQueen's objection would seem to undermine the truth of premise (ii) of the argument for the threshold. However, it is easy to see that the objection does not apply to the proposal I put forward, since indeterminacy in property instantiation \textit{does not entail} indeterminacy in existence or identity (see \cite{Wil13}). Wilson's view may well be counterintuitive, yet it is a perfectly coherent and well developed metaphysical view to which McQueen's objection fails to apply. In order to make sense of Clifton \& Monton's view, and to respond to McQueen's criticism, it is true that we have to make an explicit link between degrees of instantiation and probability---as the \textbf{Degree Link} aims to do. However, once again, such a connection does not entail, or even suggest, that existence comes in degrees.

\section{Non-Accessible Mass in the Current Debate on GRW}
\textbf{Realism} as it has been defended throughout this paper has various interesting consequences. For reasons of space, I shall leave a detailed analysis of such consequences to another occasion, yet I believe it is important to briefly mention two issues. 

First, GRW$_M$ has often been taken as a prominent example of the Primitive Ontology (PO) approach to QM \cite{All08}. According to \cite{All16}, a suitable PO (i) has to be defined in 3D space, and (ii) has to be microscopic rather than macroscopic. If met, these requirements would ground a classical reductive explanation of the behaviour of macroscopic objects as determined by the behaviour of the PO. However, very little has been said regarding whether or not the PO can be indeterminate as \textbf{Realism} towards non-accessible mass may require it to be. Presumably, the PO should be taken as determinate in order to explain the definiteness of experimental outcomes. Moreover, Bell \cite{Bel87} was quite clear that the notion of \textit{local beable} has to represent a \textit{classical} ontology, in agreement with Bohr that the experimental context has to always be described in classical terms. But while the PO approach is inspired by Bell (see \cite{All08}), I shall notice that there is arguably very little classicality in the idea of states of mass being ontologically indeterminate.  

These considerations suggest that proponents of the PO approach in the context of GRW$_M$ face a dilemma. Either they provide a better defense of \textbf{Anti-Realism}, or, if they are willing to endorse \textbf{Realism}, they could either develop a version of it that avoids indeterminacy or simply accept that the PO could be indeterminate.\footnote{As I said earlier in \textit{footnote 19}, I am of course open to the possibility that there may be ways to understand non-accessible mass states without accepting indeterminacy. I shall notice, however, that at the current stage indeterminacy is the only option that is developed and discussed enough in the literature on how to provide an ontological understanding of states of superposition (like $\ket{\psi^{\oplus}}$ of equation (5)), and this is why I decided to focus on this strategy. I would like to thank an anonymous reviewer of this journal for inviting me to mention this possibility.} Either way, this paper urges them to say more about their view. I believe that the best strategy is simply to accept that the PO could be indeterminate. And as a matter of fact, a prominent defender of the PO approach such as Tumulka seems to partially agree:

\begin{quote}
     [...] the PO does provide a picture of reality that conforms with our everyday intuition. All this is independent of whether the PO is observable (accessible) or not. Bassi and Ghirardi sometimes sound as if they did not take the matter density seriously when it is not accessible; I submit that \textit{the PO should always be taken seriously}. (\cite{Tum11}: 9, italics mine)
     \end{quote}
     
\noindent If the PO should always be taken seriously as Tumulka suggests (namely, we endorse \textbf{Realism}), and if states of non-accessible mass are, as I have argued, indeterminate states of affairs, this entails that the determinacy of the PO can no longer be a core \textit{desideratum} for such a notion.\footnote{I shall register that, on the contrary, proponents of the PO seem to take the determinacy as an essential component for their view (Valia Allori, and Vincent Lam, personal communication), despite the fact that, at least to the best of my knowledge, in papers they are never explicit about this.} 

A second major conceptual consequence of \textbf{Realism} relates more closely to the philosophical debate over indeterminacy in QM. It has been argued that in the major interpretations of QM, even if there were some indeterminacy, this can be viewed as dispensable because it would only affect a derivative level of reality (\cite{Gli17}, \cite{Che20}).\footnote{Notice that this line of reasoning is based on the assumption that any \textit{ontic indeterminacy} may be viewed as eliminable if it is not fundamental. See \cite{Bar14} for an explicit defence of this assumption, and \cite{Eva18} and \cite{Mar20} for critics. See also \cite{Mar21} for a defense of the thesis that quantum indeterminacy is emergent and yet ontological.} GRW has also been indicated as further support to this conclusion, as it is clear from this passage from Glick:

\begin{quote}
    […] consider dynamical collapse theories such as versions of GRW. The two versions of the GRW adopted by most contemporary defenders are the mass-density and flash-ontology varieties. Neither contains fundamental indeterminacy: the distribution of mass-density and the location of the flashes are both perfectly determinate. (\cite{Gli17}: 205)
\end{quote}

\noindent Although Glick maintains that on GRW$_M$ \enquote{the distribution of mass-density [is] perfectly determinate}, I have argued extensively that this is not the case unless one is willing to accept \textbf{Anti-Realism}. This consideration also indicates that, contrary to Glick's conclusion, there might be after all at least one interpretation of QM in which the indeterminacy is not entirely absent from the fundamental level.\footnote{By fundamental I mean \textit{relative to the theory} we discuss, as Glick also has it. Of course, we both agree that for the sake of the argument we should momentarily set aside the possibility of a more fundamental physical theory.} I shall register that, quite independently from the arguments for \textbf{Realism} I have given, the idea that the mass-density is \textit{perfectly determinate}, to use Glick's words, appears highly unmotivated. Take the following passage from Maudlin's recent book \cite{Mau19}:

\begin{quote}
    [I]f the square amplitude of the wavefunction assigns a weight of .25 to a configuration in which a particular electron is on the left and a weight of .75 to a configuration in which the electron is on the right, then somehow .25 of the matter of the electron is on the left, and the other .75 is on the right. Since each possible configuration assigns an exact position to each particle, the weighting of the configurations can in this way be used to define a matter distribution for each particle. \textit{The matter of the particle literally gets smeared out over space} (\cite{Mau19}: 117, italics added). 
\end{quote} 

\noindent While I do not mean to suggest that the above passage is in full agreement with the version of GRW$_M$ I have defended, it does certainly show that a pre-theoretical notion of indeterminacy is often taken as a core conceptual component of this theory. Maudlin speaks about the \textit{matter of the electron being smeared out over space}, by which he means that, accordingly, the location itself of the electron is in the right or the left only to a certain degree, and therefore it is not definite. My own version of \textbf{Realism} is not but a way of developing and refining these ideas, which also seem to reflect the most natural philosophical attitude towards GRW$_M$. And indeed, the beautiful passage from Bell which opens this paper---also referenced by Ghirardi in \cite{Bas04}---seems to deliver a similar message: the \textit{cloudiness} of electrons, understood in a realist sense as a lack of definiteness, is one of the most essential part of this theory.

\section*{Conclusions}
I have shown that the \textbf{Criterion of Accessibility} of mass in GRW$_M$ leaves the ontological meaning of the non-accessible portion of mass unexplained (\S2). I have then introduced two viable positions regarding the status of non-accessible mass, \textbf{Anti-Realism} and \textbf{Realism}, have argued that \textbf{Anti-Realism} is untenable (\S3), and that \textbf{Realism} can be coherently construed through the \textbf{Degree Link} (\S4). The form of \textbf{Realism} I have defended entails that the ontology of GRW$_M$ has to allow for the existence of indeterminate states of affairs, a result that, as I have briefly shown (\S5), has numerous important conceptual consequences, many of which are yet to be properly discussed and fully understood.

\subsection*{Acknowledgements}
For useful comments on previous versions of this paper, I wish to thank Valia Allori, Angelo Bassi, Claudio Calosi, Vincent Lam, Peter Lewis, Maria Maffei, Giuliano Torrengo, Lev Vaidman, and two anonymous referees of this journal. I also acknowledge the generous support of the French National Research Agency, under the program Investissements d'avenir (ANR-15-IDEX-02).

\end{document}